\documentclass[twocolumn]{jpsj2}

\def\runtitle{Hole-Doping Effects on a Two-dimensional Kondo Insulator
}
\def\runauthor{Yasuhiro \textsc{Saito}, 
Akihisa \textsc{Koga} and Norio \textsc{Kawakami}
}

\title{
Hole-Doping Effects on a Two-dimensional Kondo Insulator
}

\author{%
Yasuhiro Saito, Akihisa Koga and 
Norio Kawakami
}

\inst{
Department of Applied Physics, Osaka University, Suita, 
Osaka 565-0871, Japan
}

\recdate{\today}

\abst{%
We study the effects of hole doping on the two-dimensional 
Heisenberg-Kondo  model around the quantum
critical point, where the spin liquid phase (Kondo insulator) and the
 magnetically ordered phase are separated via a second-order
  phase transition. By means of the self-consistent Born
 approximation within  the bond operator formalism as well as
 the standard spin wave theory, we
 discuss dynamical properties of a doped hole. It is clarified
 that a quasi-particle state stabilized in the spin liquid phase 
is gradually obscured as the system approaches the quantum critical point.
 This is also the case for the magnetically ordered phase.
 We argue the similarity and the difference between these two cases.
}

\kword{%
Kondo Insulator, bond operator, spin wave, self-consistent Born approximation
}

\begin{document}
\sloppy
\maketitle

\section{Introduction}
Heavy fermion systems realized in rare-earth compounds 
have attracted much interest over many years. They
show a variety of striking phenomena due to strong
electron correlations. In metallic systems, the electron mass 
is renormalized to a huge value by the {\it f-f} Coulomb
interaction.\cite{review} Such renormalization effects are 
also relevant for the insulating phase, 
 leading to the Kondo insulator possessing both of the 
spin and charge gaps renormalized by electron correlations.
\cite{Kinsulator,klattice}
Since these metallic and insulating states have 
essentially the same origin of electron correlations, it is 
interesting to ask how the Kondo insulator is 
changed to the heavy fermion metallic system upon hole doping.

Another interesting aspect in heavy fermion systems 
is that some of unusual static and dynamical properties are related to
the fact that the system is located in the vicinity of a
quantum critical point.
\cite{Kuramoto,Andraka,Hertz,Millis,Moriya,Continentino,Schlottmann,Coleman,Si,Rosch,Lavagna} 
For instance, the anomalous
non-Fermi-liquid temperature dependence in the specific heat and the 
susceptibility observed in rare-earth compounds
such as $\rm CeCu_{6-x}R_x (R=Au, Ag)$,\cite{CeCu6}
$\rm CePd_2Si_2$,\cite{CePd2Si2} $\rm CeNi_2Ge_2$,\cite{CeNi2Ge2}
$\rm U_{1-x}Y_xPd_3$,\cite{UPd3} $\rm Ce_xLa_{1-x}Ru_2Si_2$\cite{LaRu2Si2}
is caused by large quantum
fluctuations reflecting the phase transition to the 
antiferromagnetically ordered phase.
 In this way, strong quantum 
fluctuations near the critical point \cite{Chakravarty}
give rise to rich and attractive phenomena in rare-earth systems.

Stimulated by these interesting topics, we consider a simplified
model of the Kondo insulator, which possesses the quantum critical 
point between the Kondo insulating phase and the antiferromagnetic phase.
 To be precise, we employ 
 the two-dimensional Heisenberg-Kondo model,
for which the charge degree of freedom is frozen at half filling, 
and the remaining spin sector is either in the spin liquid phase or
the antiferromagnetically ordered phase at zero temperature.
We investigate the effects of hole doping on the model with particular 
emphasis on the hole dynamics near
the quantum critical point. Although the effects of hole doping
into the Kondo insulator have been addressed so far,\cite{klattice}
we think that the dynamics of a doped hole has not been discussed in detail.
Also, in some respects, the present investigation is closely related to
those done for the two-dimensional Heisenberg model in connection 
with the high $T_c$ superconductors.\cite{rink,horsh,liu,frank}

This paper is organized as follows. In Sec. \ref{sec2}, we introduce 
the two-dimensional Heisenberg-Kondo model and summarize the 
basic properties of the model. 
In Sec. \ref{sec3}, we study the hole dynamics, when 
it is doped into the spin liquid phase, by combining the 
self-consistent Born approximation with the bond-operator
formalism. Similar analysis is performed for 
the antiferromagnetic phase in Sec. \ref{sec4}
by exploiting the spin wave theory. 
A brief summary is given in Sec. \ref{Summary}.

\section{Heisenberg-Kondo model}\label{sec2}

\begin{figure}[htb]
\begin{center}
\includegraphics[width=5cm]{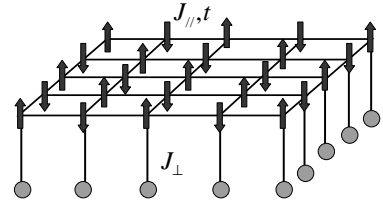}
\end{center}
\vskip -3mm   
\caption{ Heisenberg-Kondo model on the square lattice.
Spins of host electrons indicated by the arrows are antiferromagnetically 
coupled to localized $f$-spins indicated by the circles.
}
\label{fig:model}
\end{figure}

We investigate the two-dimensional Heisenberg-Kondo model 
having conduction electrons coupled to 
 localized $f$-spins. The model is schematically drawn in
Fig. \ref{fig:model}. 
The Hamiltonian reads
\begin{eqnarray}
H&=&H_{t}+H_{J},
\label{eq:hamiltonian}
\\
H_{J}&=&J_{//}\sum_{<i,j>}{\bf S}_{i}^{a}\cdot{\bf S}_{j}^{a}
+J_{\perp}\sum_{i}{\bf S}_{i}^{a}\cdot{\bf S}_{i}^{b},
\label{eq:hj}
\\
H_{t}&=&-t\sum_{<i,j>,\sigma}\hat{c}_{i,\sigma}^{\dag}\hat{c}_{j,\sigma}
+H.c. ,\label{eq:ht}
\end{eqnarray}
with $\hat{c}_{i,\sigma}=[c_{i,\sigma}(1-n_{i,
-\sigma})]$, where $c_{i,\sigma}$ 
annihilates a conduction electron 
with spin $\sigma(=\uparrow, \downarrow)$ at the $i$th site. 
The corresponding spin operator is defined by 
${\bf S}_{i}^{a}=\frac{1}{2}\sum_{\sigma,\sigma'}c_{i,\sigma}^\dag 
{\bf \tau}_{\sigma,\sigma'}c_{i,\sigma'}$, 
where ${\bf \tau}$ is the Pauli matrix.
The operator of an $s=1/2$ $f$-spin is 
denoted by ${\bf S}_{i}^{b}$.
The exchange couplings $J_{//}$ and $J_{\perp}$ are
assumed to be antiferromagnetic.
For convenience, we set hopping $t$ as unity and introduce 
the ratio $j=J_{//}/J_{\perp}$. 
At half filling, the system is in the spin-liquid insulating 
phase with both 
of the spin and charge gaps for smaller $j$, which may be a 
proper model of the Kondo insulator. We note that magnetic properties 
for $j=0$  have already been discussed.\cite{Zheng,Brenig,Assaad,Shi,Wang}
An interesting aspect in the model is that there exists a 
quantum phase transition at $j_{c}$ ($\approx 0.75$, 8-th order 
series expansion)
\cite{Matsushita} from the spin-liquid phase (Kondo insulator) to
the magnetically ordered phase as $j$ increases. Therefore,
the system realizes a quantum critical point at $j=j_c$, where
the spin gap vanishes, but a long-range order 
still does not develop. This specific condition may give rise to 
anomalous behavior of a doped hole, reflecting large 
quantum fluctuations. 

In order to discuss dynamical properties of a hole-doped
Kondo insulator in the vicinity of 
the quantum critical point, we make use of either the bond-operator
approach  \cite{Sachdev,Sigrist,Lee} or the
spin wave approach.
These methods are appropriate to discuss magnetic
properties in the spin liquid phase and the magnetically ordered
phase, respectively. 
We further exploit the self-consistent 
Born approximation(SCBA),\cite{rink,horsh,
liu,frank,jurecka}
in which the motion of hole can be treated as 
a scattering problem in the host spin system.
In the following, we separately discuss dynamical properties 
of a doped hole in the spin liquid phase and the magnetically ordered phase.

%
\section{Hole doping in the spin-liquid phase}\label{sec3}
%

We begin with dynamical properties of a doped hole in the
spin liquid phase (Kondo insulator) having the spin gap for excitations.
For this purpose, we use the bond-operator representation, which is
useful to deal with the spin-singlet state.

\subsection{bond-operator representation}\label{sec3a}

When the Kondo coupling between the host spins and localized spins
is fairly strong, the dimer singlet is formed at each site, which
may be a good starting point to describe the spin liquid phase.
This condition allows us to
 introduce six kinds of the bond operators, which are 
defined at each site as, 
\begin{eqnarray}
s_{n}^{\dag}|0\rangle&=&\frac{1}{\sqrt{2}}(|\uparrow \downarrow \rangle-
|\downarrow \uparrow \rangle),
\nonumber\\
t_{x,n}^{\dag}|0\rangle&=&\frac{-1}{\sqrt{2}}(|\uparrow \uparrow \rangle-
|\downarrow \downarrow \rangle),
\nonumber\\
t_{y,n}^{\dag}|0\rangle&=&\frac{i}{\sqrt{2}}(|\uparrow \uparrow \rangle+
|\downarrow \downarrow \rangle),
\nonumber\\
t_{z,n}^{\dag}|0\rangle&=&\frac{1}{\sqrt{2}}(|\uparrow \downarrow \rangle+
|\downarrow \uparrow \rangle),
\nonumber\\
a_{n,\uparrow}^{\dag}|0\rangle&=&|\circ 
\uparrow\rangle,\nonumber\\
a_{n,\downarrow}^{\dag}|0\rangle&=&|\circ 
\downarrow\rangle,
\label{eq:bond}
\end{eqnarray}
where ket states are specified by the configuration of 
a conduction electron and an $f$-electron [$\circ$ 
represents an unoccupied (hole) site].
Here, the operators $s^{\dag}$ and $t_\alpha^{\dag} (\alpha=x, y, z)$ obey
the bosonic commutation 
relation, while the operator $a^\dag$ the fermionic 
anticommutation relation.
To restrict the physical states to either singlet, triplet or
pseudofermion, the above operators are subjected to the constraint,
\begin{eqnarray}
s_{n}^{\dag}s_{n}+\sum_{\alpha}t_{\alpha,n}^{\dag}t_{\alpha,n}+
\sum_{\sigma}a_{n,\sigma}^{\dag}a_{n,\sigma}=1.
\label{eq:limit}
\end{eqnarray}    

In the spin liquid phase, 
the boson $s$ is condensed, which results in the finite value of 
 $<s_n>=\bar{s}$. 
Then the spin Hamiltonian  $H_J$ is diagonalized in the Fourier
space as,
\begin{eqnarray}
H_{J} &= & \sum_{{\bf k},\alpha}\omega_{{\bf k}}\gamma_{\alpha,{\bf k}}
^{\dag}\gamma_{\alpha,{\bf k}}+{\rm const.},\\
\label{eq:omega}
\omega_{{\bf k}}& = &J_{\perp}\sqrt{1+2e_{{\bf k}}},
\nonumber
\label{eq:hgp}
\end{eqnarray}
 where $e_{{\bf k}}
=j\bar{s}^{2} Q({\bf k})$ with 
$Q({\bf k}) = \frac{1}{2}(\cos{k_{x}} + \cos{k_{y}})$. 
Here, we have introduced the normal-mode operator 
$\gamma_{\alpha,{\bf k}}^{\dag}$ 
related to $t_{\alpha,{\bf k}}^{\dag}$ via the Bogoliubov transformation,
\begin{eqnarray}
t_{\alpha,{\bf k}}^{\dag}=u_{{\bf k}}
\gamma_{\alpha,{\bf k}}^{\dag}+v_{{\bf k}}\gamma_{\alpha,-{\bf k}}, 
\label{eq:bogo}
\end{eqnarray}
where $v_{{\bf k}}^{2}=1/2 \{J_{\perp}(1+e_{{\bf k}})/\omega_{{\bf k}}-1\}$, 
$u_{{\bf k}}^{2}=1/2 \{J_{\perp}(1+e_{{\bf k}})/\omega_{{\bf k}}+1\}$.


When a hole is doped into the system, 
the condensate density of the singlet is determined by 
the hard-core constraint eq. (\ref{eq:limit}) as
\begin{eqnarray}
\bar{s}=1-\frac{3}{N}\sum_{{\bf q}}v_{{\bf q}}^{2}.
\label{eq:sing}
\end{eqnarray}    
In the bond-operator representation, the operator 
for a physical electron is
written by two bond operators as, 
%
\begin{eqnarray}
\hat{c}_{n,\sigma}=\frac{1}{\sqrt{2}}&[&a_{n,\bar{\sigma}}^{\dag}
(p_{\sigma}s_{n}+t_{z,n})
\nonumber\\ 
&+&a_{n,\sigma}^{\dag}(p_{\bar{\sigma}}
t_{x,n}+it_{y,n})],
\label{eq:hole-creation}
\end{eqnarray}    
%
where $p_{\sigma}=+(-)$, $\bar{\sigma}=\downarrow(\uparrow)$ for 
$\sigma=\uparrow(\downarrow)$. 
We thus obtain the Hamiltonian for electron hopping,
\begin{eqnarray}
H_{t}&=&\frac{t}{2}\bar{s}^{2}\sum_{<i,j>,\sigma}a_{i,\sigma}^{\dag} 
a_{j,\sigma}
-t\bar{s}\sum_{<i,j>}({\bf t}_{j}^{\dag}{\bf T}_{i,j}
+{\bf t}_{i}^{\dag}{\bf T}_{j,i})
\nonumber\\
&&+\frac{t}{2}\sum_{<i,j>,\sigma}{\bf t}_{i}^{\dag}{\bf t}_{j}
a_{j,\sigma}^{\dag} 
a_{i,\sigma}
-t\sum_{<i,j>}i({\bf t}_{i}^{\dag}\times {\bf t}_{j}){\bf T}_{j,i}
\nonumber\\
&&+ H.c.
\label{eq:hami-1}
\end{eqnarray}    
with ${\bf t}_{i}^{\dag}=(t_{x,i}^{\dag},t_{y,i}^{\dag},t_{z,i}^{\dag})$
and
\begin{eqnarray}
{\bf T}_{m,n}=\frac{1}{2}\sum_{\sigma_{1},\sigma_{2}}a_{m,\sigma_{1}}
^{\dag}{\bf \tau}_{\sigma_{1},\sigma_{2}}a_{n,\sigma_{2}},
\label{eq:new-spin}
\end{eqnarray}    
where $({\bf T}_{m,n})^{\dag}={\bf T}_{n,m}$. 
This Hamiltonian is spin rotationally invariant, 
reflecting the fact that the system has no magnetic order. 

\subsection{Green function}

\begin{figure}[htb]
\begin{center}
\includegraphics[width=5cm]{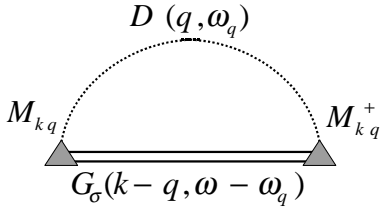}
\end{center}
\vskip -3mm
\caption{Self-energy of a pseudofermion in the SCBA. 
The triangle represents the scattering matrix of a triplet boson. 
}
\label{fig:SCBA}
\end{figure}

To discuss the dynamics of a doped hole, we define the retarded Green
functions for a pseudofermion and a physical electron as
\begin{eqnarray}
G_{\sigma}({\bf k},t)&=&-i\Theta(t)<D|\{a_{{\bf k},\sigma}(t),
a_{{\bf k},\sigma}^{\dag}\}|D>,
\label{eq:green-1}
\\
G_{\sigma}^{c}({\bf k},t)&=&-i\Theta(t)<D|\{\hat{c}_{{\bf k},\sigma}(t),
\hat{c}_{{\bf k},\sigma}^{\dag}\}|D>,
\label{eq:green-2}
\end{eqnarray}    
where $|D>$ represents the spin-singlet ground state.
For small $j$, 
renormalization effects due to the two 
triplet vertices in eq. (\ref{eq:hami-1}) are of minor importance,
so that we will discard them in the following. 
We then arrive at the Hamiltonian for a doped hole as 
\begin{eqnarray}
H_{0}&=&2t\bar{s}^{2}\sum_{{\bf k}}Q({\bf k})a_{{\bf k},\sigma}^{\dag}
a_{{\bf k},\sigma},
\label{eq:hami-2}
\\
V&=&-\frac{1}{\sqrt{N}}\sum_{{\bf k},{\bf q},\sigma_{1},\sigma_{2}}
\{
M_{{\bf k}{\bf q}}{\bf \gamma}_{{\bf q}}^{\dag}a_{{\bf k}-{\bf q},
\sigma_{1}}^{\dag}{\bf \tau}_{\sigma_{1},\sigma_{2}}a_{{\bf k},\sigma_{2}}
\nonumber\\
&&+M_{{\bf k}{\bf q}}{\bf \gamma}_{-{\bf q}}a_{{\bf k}-{\bf q},
\sigma_{1}}^{\dag}{\bf \tau}_{\sigma_{1},\sigma_{2}}a_{{\bf k},\sigma_{2}}
\},
\label{eq:hami-3}
\end{eqnarray}
where the element
 $M$ in the perturbation term $V$ 
 represents the scattering of a triplet boson, 
\begin{eqnarray}
M_{{\bf k}{\bf q}}=t\bar{s}\{Q({\bf k})+Q({\bf k}-{\bf q})\}
(u_{{\bf q}}+v_{{\bf q}}).
\label{eq:mkq}
\end{eqnarray}
%
We wish to note that the SCBA is useful to deal with the Hamiltonian eqs. 
(\ref{eq:hami-2}) and (\ref{eq:hami-3}).
In terms of the bare Green functions for a pseudofermion and
a triplet-boson,
\begin{eqnarray}
G_{\sigma}^{0}({\bf k},\omega)=\frac{1}{\omega-2t\bar{s}^{2}
Q({\bf k})},
\label{eq:bare}
\end{eqnarray}
\begin{eqnarray}
D_{\alpha}({\bf k},\omega)=\frac{1}{\omega-\omega_{{\bf k}}},
\label{eq:tgreen}
\end{eqnarray}
we can evaluate the self-energy of a pseudofermion
in the SCBA, which is 
diagramatically shown in Fig. \ref{fig:SCBA}. 
The Green function defined by $G_{\sigma}({\bf k},\omega)=
[G_{\sigma}^{0}({\bf k},\omega)^{-1}
-\Sigma_{\sigma}({\bf k},\omega)]^{-1}$
is determined via the self-consistent equation,
%
\begin{eqnarray}
{\bf\Sigma}_{\sigma}({\bf k},\omega)
&=&\frac{3}{N}\sum_{{\bf q}}
M_{{\bf k}{\bf q}}^{2}[G_{\sigma}^{0}({\bf k}-{\bf q},\omega-
\omega_{q})^{-1}
\nonumber\\
&&-{\bf\Sigma}_{\sigma}({\bf k}-{\bf q},\omega-\omega_{{\bf q}})
]^{-1},
\label{eq:s-energy-1}
\end{eqnarray}    
where $\omega=\omega+i\eta$ with 
$\eta \rightarrow 0^{+}$. 

To obtain the physical Green function of an electron,
we introduce the elements as
\begin{eqnarray}
\alpha_{{\bf k},\sigma}&=&
s \langle D 
     |a_{{\bf k},\bar{\sigma}}\hat{c}_{{\bf k},\sigma}
   | D \rangle
=\frac{\pm s}{\sqrt{2}},
\nonumber\\
\beta_{{\bf k}{\bf q},\sigma}&=&\langle D 
     |(\gamma_{{\bf q}}a_{{\bf k}-{\bf q}})_{(1/2)(\mp 1/2)}
     \hat{c}_{{\bf k},\sigma}
     |D\rangle
\nonumber\\
&=&\pm \sqrt{\frac{3}{2N}}v_{q},
\label{eq:alpha-beta}
\end{eqnarray}    
which relate the
Green function of a pseudofermion to that of a real electron.
When $j<j_{c}$, the density of triplets is small in the ground state, 
i.e., $0<v_{{\bf q}}\ll 1$.
Then it is legitimate to
 use a perturbative evaluation of two-particle propagator 
within the random-phase approximation (RPA) \cite{jurecka}.
This gives us the Green function,
\begin{eqnarray}
&&G_{\sigma}^{c}(k,\omega)=\alpha_{k,\sigma}G_{\sigma}(k,\omega)\alpha_{k,\sigma}
\nonumber\\
&&+\sum_{q}\beta_{kq,\sigma}G_{\sigma}(k-q,\omega-\omega_{q})\beta_{kq,\sigma}
\nonumber\\
&&+\left(\alpha_{k,\sigma}G_{\sigma}(k,\omega), 
\sqrt{\frac{3}{N}}\sum_{q}\beta_{kq,\sigma}G_{\sigma}(k-q,\omega-\omega_{q})
 M_{kq}^{\dag} \right)
\nonumber\\
&&\times \left(\begin{array}{cc}
1 & -\frac{3}{N}\sum_{q}M_{kq}G_{\sigma}(k-q,\omega-\omega_{q})M_{kq}^{\dag}
\\
-G_{\sigma}(k,\omega) &1
\end{array}
\right)^{-1}
\nonumber\\
&& \times \left( \begin{array}{r}
\sqrt{\frac{3}{N}}\sum_{q}M_{kq}G_{\sigma}(k-q,\omega-\omega_{q})\beta_{kq,\sigma} \\
G_{\sigma}(k,\omega)\alpha_{k,\sigma}
\end{array}
\right).
\label{eq:rpa}
\end{eqnarray}
This completes the formulation of the SCBA for a hole doped 
in the spin liquid phase.

\subsection{numerical results}

\begin{figure}[htb]
\begin{center}
\includegraphics[width=6cm]{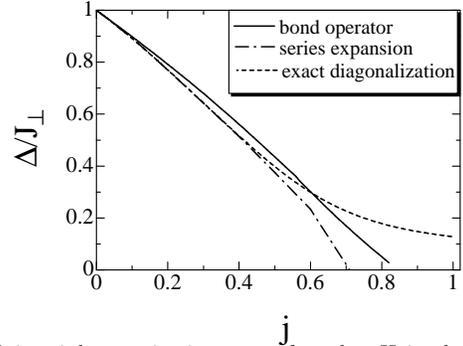}
\end{center}
\vskip -5mm	   
\caption{Spin-triplet excitation gap for the Heisenberg-Kondo model 
at half filing. 
The solid, dash-dotted, and broken lines represent
the spin gap computed by the bond-operator method, the series expansion
method (9-th order), and the exact diagonalization 
($3\times 4$ system).
}
\label{fig:phase-d}
\end{figure}

Before discussing the nature of the hole-propagator, we check 
how well our bond-operator approach works at half filling.
In Fig. \ref{fig:phase-d}, we show the spin-triplet excitation gap
calculated by several different methods.
The phase transition point was evaluated approximately as 
 $j_c \simeq 0.75$,\cite{Matsushita} 
 so that we can say that the bond-operator approach may 
 describe the spin-liquid 
phase rather well in a wide range of the parameters.

\begin{figure}[htb]
\begin{center}
\includegraphics[width=7.5cm]{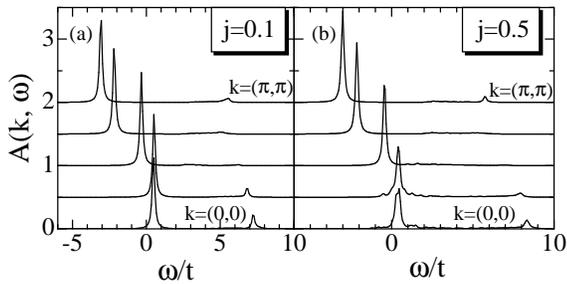}
\end{center}
\vskip -3mm
\caption{Spectral function of a doped hole
in the spin liquid phase for $J_{\perp}=5$, 
$j=0.1$ (a) and 
$j=0.5$ (b). 
The momentum of each spectral function is chosen as
${\bf k}= (0,0), (\pi/4,\pi/4), (\pi/2,\pi/2), (3\pi/4, 3\pi/4),
(\pi,\pi)$ from the bottom to the top.
}
\label{fig:dimer-1}
\end{figure}

\begin{figure}[htb]
\begin{center}
\includegraphics[width=5.5cm]{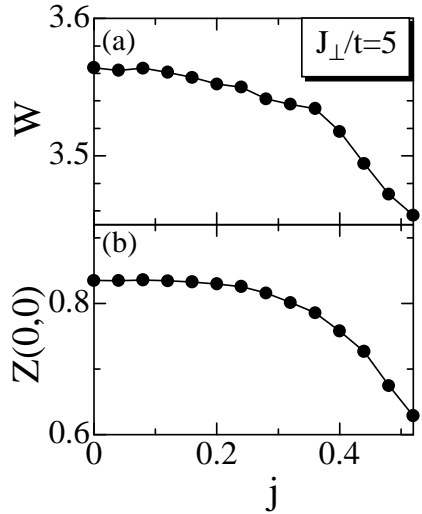}
\end{center}
\caption{
Plots of the 
width $W$ of the quasi-particle band and
the renormalization factor $Z$ for the spin-liquid phase.
}
\label{fig:dimer-n}
\end{figure}

Let us now turn to the hole-doped case. From now on, we set $\eta=0.1$. 
Shown in Fig. \ref{fig:dimer-1} is the spectral function of 
a doped hole computed by the SCBA. 
When $J_{\perp}=5$ and $j=0.5$, 
we find a dominant dispersive peak in the spectral function,
which originates from the first term in the Hamiltonian 
(\ref{eq:hami-1}), indicating the 
coherent motion of a doped hole without 
triplet excitations. 
Namely, the low-energy edge of the spectrum 
determined by the poles of the
Green function on the real axis remains separated by 
gap from the continuum of multimagnon shake off. 
The corresponding dispersion of a quasiparticle 
is given by
\begin{eqnarray}
\epsilon_{{\bf k}}=2t\bar{s}^{2}Q({\bf k})-\frac{3}{N}\sum_{{\bf q}}
\frac{M_{{\bf k}{\bf q}}^{2}}{\omega_{\bf q}}.
\label{eq:dispersion-1}
\end{eqnarray}
\begin{figure}[htb]
\begin{center}
\includegraphics[width=6cm]{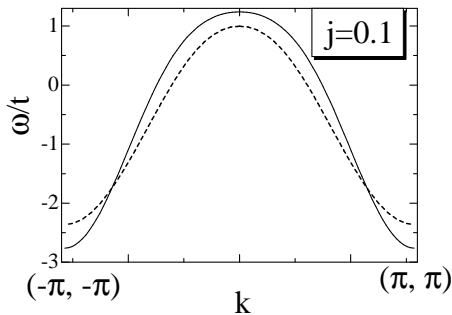}
\end{center}
\caption{Dispersion relation of a doped hole 
for $J_{\perp}=5$ and $j=0.1$. 
The solid (broken) line is obtained by the SCBA (series expansion 
of the 7th order).
}
\label{fig:dimer-2}
\end{figure}
Shown in Fig. \ref{fig:dimer-2} is the quasi-particle dispersion
thus obtained. 
The comparison of this with the spectral function 
in Fig. \ref{fig:dimer-1} confirms that the sharp
peak is indeed developed along the dispersion curve.
This implies that the quasi-particle state is quite stable in this 
parameter region.
We also calculate the dispersion of a hole by
means of the series expansion,\cite{Gelfand} in which the
excitation energy is expanded 
in $j$ and $t$ up to the seventh order. The results are shown 
in Fig. \ref{fig:dimer-2} by the broken line.
We find that the results obtained by the SCBA are consistent with those
obtained by the series expansion. 
Increasing the interdimer coupling $J_{//}$, antiferromagnetic
correlations are gradually enhanced and the spin gap is decreased.
Then a doped hole suffers from the scattering by low-lying
spin excitations, making the peak structure on the 
dispersion curve somehow broadened.
This tendency is clearly observed in 
the peak around ${\bf k}=(0,0)$ 
in Fig. \ref{fig:dimer-1}(b). 
When $J_{\perp}=1$ and $j=0.5$, 
\begin{figure}[htb]
\begin{center}
\includegraphics[width=7.5cm]{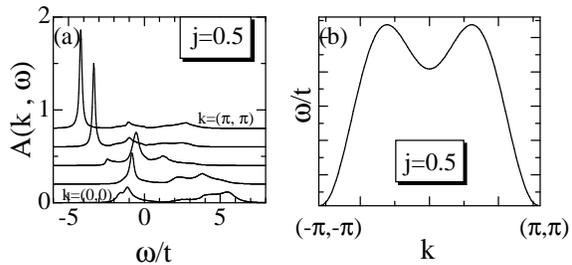}
\end{center}
\caption{
Spectral function of a doped hole (a) and the corresponding 
dispersion relation (b) for $J_{\perp}=1$ and $j=0.5$.
The momenta are chosen as in Fig. \ref{fig:dimer-1}.
}
\label{fig:dimer-4}
\end{figure}
the spin gap becomes small.
In this case, the quasiparticle peak around ${\bf k}= (0,0)$
is completely smeared, as shown in Fig. \ref{fig:dimer-4}. 
The large portion of the corresponding weight is shifted to 
 the incoherent parts in higher energy region. The dispersion 
 relation obtained approximately possesses a dip structure 
 around the origin, as seen in Fig. \ref{fig:dimer-4}(b).
 However, this structure may not be sensible,
 since the quasi-particle picture does not hold in this
momentum region. We note that even in this case, the spectrum 
 around ${\bf k}=(\pi, \pi)$ still forms a rather sharp 
 structure, implying that the hole motion with this momentum
is not affected so much by spin excitations. 
For reference we show 
the width $W$ of the quasi-particle band 
and the renormalization factor $Z$
 of the ${\bf k}=(0,0)$ state in Fig. \ref{fig:dimer-n}.
 The latter quantity is defined by 
 \begin{eqnarray}
Z({\bf k})=\left (1-\frac{\partial \Sigma({\bf k}, \omega)}
{\partial \omega} \right )^{-1}|_{\omega=\epsilon_{{\bf k}}},
\label{eq:weight}
\end{eqnarray}
which corresponds to the weight of the quasi-particle state.
It is seen that the band width becomes small as the system approaches
the quantum critical point. At the same time,
 the weight of the quasi-particle decreases
monotonically near the critical point, being consistent with 
the disappearance of the well-defined 
quasi-particle state.
 We have so far discussed spectral properties of a doped hole in 
the spin liquid phase, and shown that well-defined quasi-particle states
in the $j \ll j_c$ are gradually obscured when the system
approaches the quantum critical point $j=j_c$. In particular, 
the spectrum around ${\bf k}= (0,0)$
is affected considerably. It is to be noted here
that these characteristic properties may not be specific to 
the present model, but more generically hold for hole-doped systems 
in the spin liquid phase. 

\begin{figure}[htb]
\begin{center}
\includegraphics[width=7cm]{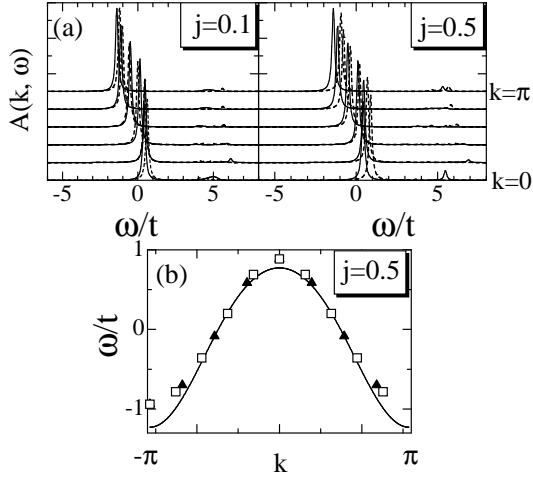}
\end{center}
\caption{(a): Spectral function of a doped hole in the one-dimensional 
Kondo insulator 
at $J_{\perp}=5$. 
The solid line is obtained by the SCBA and the dashed line by
the exact diagonalization for the system of 10 sites. The momentum is 
varied as $ k=0, \pi/4, \pi/2, 3\pi/4, \pi$. 
(b): The solid line is the dispersion relation obtained by the SCBA at 
$J_{\perp}=5$. 
The triangles and the squares represent
the results obtained by the exact diagonalization for the system
of 8 and 10 sites, respectively.
}
\label{fig:dimer-5}
\end{figure}

Before closing the section, we wish to make closer comparison of the 
present SCBA with the exact diagonalization calculation
by taking the one-dimensional model.
In Fig. \ref{fig:dimer-5}, we show the spectral function and the dispersion 
relation for the one-dimensional Heisenberg-Kondo model.
It is found that the results of the SCBA agree fairly well with 
those of the exact diagonalization, both of which show the
well-defined quasi-particle state. In this way, even in one dimension, 
 the existence of the spin gap plays an important role
in stabilizing the quasi-particle state in the spin liquid phase.

\section{Hole doping in the antiferromagnetic phase}\label{sec4}

We next discuss the effects of hole doping on 
the antiferromagnetically ordered phase $ j > j_c $. In this case, 
excitations in the host spin system may be well 
described by the spin wave theory, which we will 
use in the following discussions.
 
 \subsection{spin-wave theory}
 
We exploit the Holstein-Primakoff transformation, 
 which expresses the spin-1/2 operators 
${\bf S}_{i}^{a}$ and 
${\bf S}_{i}^{b}$ in terms
of the Bose operators $b$ and $d$ as,
\begin{eqnarray}
S_{i}^{a,+}&=&\sqrt{(1-b_{i}^{\dag}b_{i})}b_{i}
\nonumber\\
S_{i}^{a,-}&=&b_{i}^{\dag}\sqrt{(1-b_{i}^{\dag}b_{i})}
\nonumber\\
S_{i}^{a,z}&=&\frac{1}{2}-b_{i}^{\dag}b_{i}
\nonumber\\
S_{i}^{b,+}&=&d_{i}^{\dag}\sqrt{(1-d_{i}^{\dag}d_{i})}
\nonumber\\
S_{i}^{b,-}&=&\sqrt{(1-d_{i}^{\dag}d_{i})}d_{i}
\nonumber\\
S_{i}^{b,z}&=&-\frac{1}{2}+d_{i}^{\dag}d_{i}\label{eq:sw}
\end{eqnarray}
for each sublattice. 
By performing the Fourier transformation and then applying
the standard Bogoliubov transformation, 
the Hamiltonian $H_J$ for the spin sector can be diagonalized
as
\begin{eqnarray}
H_{J}=\sum_{{\bf k}}(\omega_{1{\bf k}}\alpha_{{\bf k}}^{\dag}
\alpha_{{\bf k}}+\omega_{2{\bf k}}\beta_{{\bf k}}^{\dag}
\beta_{{\bf k}})+{\rm const.},
\label{eq:order-6}
\end{eqnarray}
with $\omega_{1{\bf k}}=4J_{//}\sqrt
{B_{{\bf k}}-A_{{\bf k}}}$ and $\omega_{2{\bf k}}=4J_{//}\sqrt
{B_{{\bf k}}+A_{{\bf k}}}$, where 
$A_{{\bf k}}=1/16[4+4j^{-1}+Q({\bf k})^{2}(j^{-2}-4j^{-1}-8)
+4Q({\bf k})^{4}]^{1/2}$, $B_{{\bf k}}=1/8(1-Q({\bf k})^{2}+j^{-1}/2)$. 
Here, we have ignored the hard-core 
constraint in eq. (\ref{eq:sw}) 
 since it has little effect on low-energy properties for
small values of $j^{-1}$. The Bogoliubov transformation
from the original bosons $(b_{{\bf k}},d_{{\bf k}})$  
to the normal-mode bosons $(\alpha_{{\bf k}}, \beta_{{\bf k}})$
takes the form,
\begin{eqnarray}
\left( \begin{array}{c}
b_{{\bf k}} \\
d_{{\bf k}} \\
b_{-{\bf k}}^{\dag} \\
d_{-{\bf k}}^{\dag}
\end{array}
\right)=\left( \begin{array}{cc}
\Gamma_{{\bf k}} &-\Theta_{{\bf k}} \\
-\Theta_{{\bf k}} &\Gamma_{{\bf k}}
\end{array}
\right)\left( \begin{array}{c}
\alpha_{{\bf k}} \\
\beta_{{\bf k}} \\
\alpha_{-{\bf k}}^{\dag} \\
\beta_{-{\bf k}}^{\dag}
\end{array}
\right),
\label{eq:oeder-4}
\end{eqnarray}
where 
\begin{eqnarray}
\Gamma_{{\bf k}}&=&\left( \begin{array}{cc}
R_{{\bf k},+} &R_{{\bf k},-} \\
I_{{\bf k},+} &I_{{\bf k},-} 
\end{array}
\right)
\nonumber\\
\Theta_{{\bf k}}&=&\left( \begin{array}{cc}
L_{{\bf k},+} &L_{{\bf k},-} \\
1 &1 
\end{array}
\right)
\label{eq:oeder-5}
\end{eqnarray}
and
\begin{eqnarray}
R_{{\bf k},\pm}&=&\frac{8E_{{\bf k},\pm}}{j^{-1}p_{{\bf k},\pm}}
\nonumber\\
I_{{\bf k},\pm}&=&\frac{2}{Q({\bf k})F_{{\bf k},\pm}}
\{-\frac{j^{-2}}{64}+E_{{\bf k},\pm}(\frac{1}{2}+F_{{\bf k},\pm})\}/p
_{{\bf k},\pm}
\nonumber\\
L_{{\bf k},\pm}&=&\frac{1}{Q({\bf k})j^{-1}}[16\{-\frac{j^{-2}}{64}
+E_{{\bf k},\pm}(\frac{1}{2}+F_{{\bf k},\pm})\}]/p_{{\bf k},\pm}
\end{eqnarray}
with $p_{{\bf k},\pm}^{2}=R_{{\bf k},\pm}^{2}
+I_{{\bf k},\pm}^{2}
-L_{{\bf k},\pm}^{2}-1$. 
Here, $E_{{\bf k},\pm}=j^{-1}/8+\sqrt{B_{{\bf k}} \mp A_{{\bf k}}}$, 
$F_{{\bf k},\pm}=j^{-1}/8-\sqrt{B_{{\bf k}} \mp A_{{\bf k}}}$.

\begin{figure}[htb]
\begin{center}
\includegraphics[width=6cm]{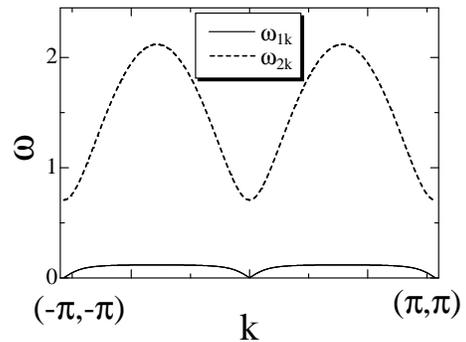}
\end{center}
\caption{Spin wave spectra at $J_{//}=1,j^{-1}=0.25$ 
along the line in the momentum space
 from ${\bf k}=(-\pi,-\pi)$ to ${\bf k}=(\pi,\pi)$. 
The solid line indicates $\omega_{1{\bf k}}$ and the dashed 
line $\omega_{2{\bf k}}$.
}
\label{fig:order-1}
\end{figure}

\begin{figure}[htb]
\begin{center}
\includegraphics[width=6cm]{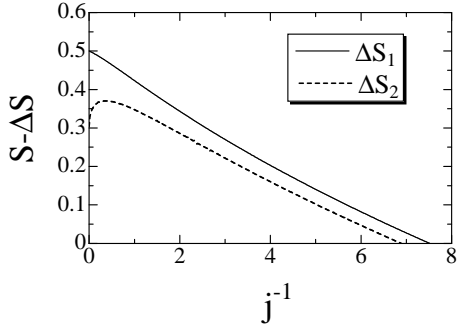}
\end{center}
\caption{Staggered magnetizations 
$S-\Delta S_{1}$ (solid line) and $S-\Delta S_{2}$
(dashed line) as a function of  $j^{-1}$.
}
\label{fig:order-2}
\end{figure}

We show the dispersion relation obtained for the undoped case
in Fig. \ref{fig:order-1},
in which gapless excitations in the low-energy region $\omega_1$ and 
 gapful excitations with a large dispersion $\omega_2$ appear.
It is easily seen that $\omega_1$ is mainly
contributed by localized spins, 
while $\omega_2$ by conduction electrons in this parameter region. 
To discuss the stability of the magnetically ordered ground
state, we calculate the spin deviation $\Delta S_{1}$ for 
$f$-spins and $\Delta S_{2}$ for conduction electrons,
\begin{eqnarray}
\Delta S_{1}=<d_{{\bf k}}^{\dag}d_{{\bf k}}>=\frac{1}{N}\sum_{{\bf k}}
(p_{{\bf k},+}^{2}+p_{{\bf k},-}^{2}),
\nonumber\\
\Delta S_{2}=<b_{{\bf k}}^{\dag}b_{{\bf k}}>=\frac{1}{N}\sum_{{\bf k}}
(L_{{\bf k},+}^{2}+L_{{\bf k},-}^{2}).
\label{eq:oeder-7}
\end{eqnarray}
In Fig. \ref{fig:order-2}, we show the staggered magnetization 
$S-\Delta S$ as a function of $j^{-1}$.
When $J_{\perp}=0$ $ (j^{-1}=0)$, the system is reduced to the 
Heisenberg antiferromagnet on the square lattice,\cite{Sakai}
which is completely separated 
from free $f$-spins $S^{b}$. 
The introduction of the exchange coupling $J_{\perp}$ enhances 
spin correlations between conduction 
electrons and localized spins.
Then the magnetization of conduction electrons 
once increases and has a maximum value around 
$j^{-1} \approx 0.37$. 
Further increase of the exchange coupling enhances the dimer
correlation, and thereby suppresses the magnetization. 
Note that the ordered state is stabilized
up to the critical value $j_{c}^{-1}\approx 6.97$ for conduction
electrons, but  $j_{c}^{-1}\approx 7.60$ for localized spins.
This pathological result means that beyond $j_{c}^{-1}\approx 6.97$
our spin-wave theory may break down.
Since the critical value $j_{c}^{-1}\approx 6.97$
is slightly larger than that 
obtained by the series expansion, we believe that
our spin wave theory can properly describe 
the ordered state except for the region very close to the
phase transition point.

\subsection{Green function}

Let us now turn to the hole-doped case.
In the spin-wave approximation, the Hamiltonian 
for electron hopping reads
\begin{eqnarray}
H_{t}&=&-t\sum_{<i,j>}\{h_{i}^{\dag}h_{j}[b_{j}^{\dag}(1-b_{i}^{\dag}
b_{i})
\nonumber\\
&&+(1-b_{j}^{\dag}b_{j})b_{i}]+H.c.\},
\label{eq:oeder-1}
\end{eqnarray}
where $h_i (=c_{i\sigma}^{\dag})$ is 
the annihilation operator of a hole. 
This is rewritten in terms of the normal modes as,
\begin{eqnarray}
H_{t}&=&-\frac{4t}{\sqrt{N}}\sum_{{\bf k},{\bf q}}h_{{\bf k}}^{\dag}
h_{{\bf k}-{\bf q}}[(R_{{\bf q},+}\gamma_{{\bf k}-{\bf q}}-
L_{{\bf q},+}\gamma_{{\bf k}})
(\alpha_{{\bf q}}+\alpha_{-{\bf q}}^{\dag})
\nonumber\\
&&+(R_{{\bf q},-}\gamma_{{\bf k}-{\bf q}}-L_{{\bf q},-}\gamma_{{\bf k}})
(\beta_{{\bf q}}+\beta_{-{\bf q}}^{\dag})].
\end{eqnarray}
We introduce the following Green function of a doped hole,
\begin{eqnarray}
G({\bf k},t)=-i\langle \psi|T[h_{{\bf k}}(t)h_{{\bf k}}^{\dag}(0)]|\psi
\rangle,
\label{eq:order-8}
\end{eqnarray}
where $|\psi \rangle$ is the ground state of the model
and $T$ is a time-ordering operator. 
When a doped hole hops in the lattice, 
the ordered state is perturbed and magnons are excited.
Note that the motion of a hole scatters two types of magnons 
specified by $\alpha_{{\bf k}}$ and $\beta_{{\bf k}}$. 
By exploiting the SCBA for magnon scattering, we have the self-energy
of the hole Green function as 
\begin{eqnarray}
\Sigma({\bf k},\omega)&=&\Sigma^{(1)}({\bf k},\omega)
+\Sigma^{(2)}({\bf k},\omega)\\
\Sigma^{(1)}({\bf k},\omega)&=&\frac{16t^{2}}{N}\sum_{{\bf q}}(R_{{\bf q},+}
\gamma_{{\bf k}-{\bf q}}-L_{{\bf q},+}
\gamma_{{\bf k}})^{2}G({\bf k}-{\bf q},
\omega-\omega_{1,{\bf q}})
\nonumber\\
\Sigma^{(2)}({\bf k},\omega)&=
&\frac{16t^{2}}{N}\sum_{{\bf q}}(R_{{\bf q},-}\gamma_{
{\bf k}-{\bf q}}-L_{{\bf q},-}\gamma_{{\bf k}})^{2}G({\bf k}-{\bf q},
\omega-\omega_{2,{\bf q}}).
\nonumber\\
\label{eq:order-9}
\end{eqnarray}

When $J_{//}\gg J_{\perp}$ and $t$, 
one obtains the dispersion relation of a hole 
analytically as,
\begin{eqnarray}
\epsilon_{{\bf k}}&=&-\xi_{1{\bf k}}-\xi_{2{\bf k}}
\nonumber\\
\xi_{1{\bf k}}&=&
\frac{16t^{2}}{N}\sum_{{\bf q}} \frac{(R_{{\bf q},+}\gamma_{
{\bf k}-{\bf q}}-L_{{\bf q},+}\gamma_{{\bf k}})^{2}}{\omega_{1{\bf q}}}
\nonumber\\
\xi_{2{\bf k}}&=&\frac{16t^{2}}{N}\sum_{{\bf q}}
\frac{(R_{{\bf q},-}\gamma_{
{\bf k}-{\bf q}}-L_{{\bf q},-}\gamma_{{\bf k}})^{2}}
{\omega_{2{\bf q}}}.
\label{eq:order-11}
\end{eqnarray}

\subsection{numerical results}

\begin{figure}[htb]
\begin{center}
\includegraphics[width=7cm]{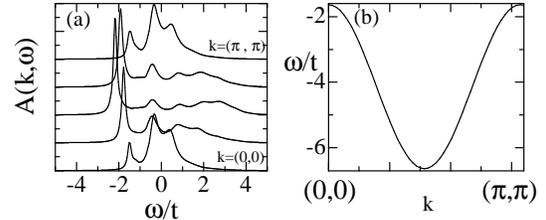}
\end{center}
\caption{(a) Spectral function of a doped hole. 
The momenta are chosen as in Fig. \ref{fig:dimer-1}.
(b) Dispersion relation given by 
Eq. (\ref{eq:order-11}): $j=0.5$, 
$J_{\perp}=10^{-5}$.
}
\label{fig:order-4}
\end{figure}

We show the spectral function in Fig. \ref{fig:order-4}(a). 
When a hole is doped into the system, 
the peak structure develops along the dispersion 
curve shown 
in Fig. \ref{fig:order-4}(b), being consistent
with those discussed in the doped Heisenberg 
antiferromagnet on the square lattice.
\cite{rink,horsh,liu,frank} 
In particular, the peak is 
rather sharp in the low-energy region near ${\bf k}\approx 
(\pi/2,\pi/2)$. 
These results imply that the quasi-particle state defined by
eq. (\ref{eq:order-11}) is stable in this parameter region.

\begin{figure}[htb]
\begin{center}
\includegraphics[width=7cm]{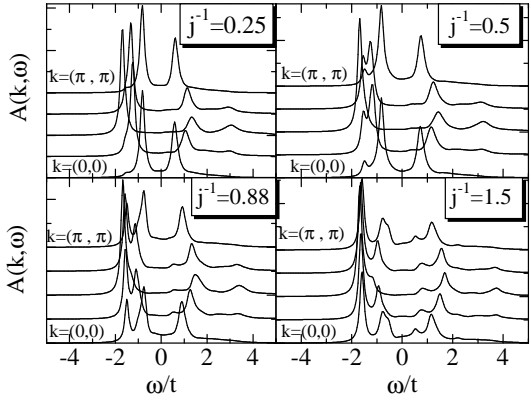}
\end{center}
\caption{Spectral function of a doped hole
as a function of 
$\omega$ for 
$J_{//}$=0.8.
The momenta are chosen as in Fig. \ref{fig:dimer-1}.
}
\label{fig:order-5}
\end{figure}

Shown in Fig. \ref{fig:order-5} is 
 the spectral function $A({\bf k},\omega)$ 
when the exchange coupling $J_\perp$ is varied.
\begin{figure}[htb]
\begin{center}
\includegraphics[width=7cm]{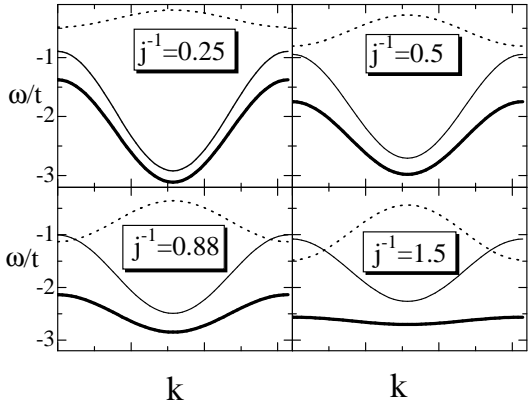}
\end{center}
\caption{
Dispersion relation (bold line) 
of a doped hole in the antiferromagnetic phase for
$J_{//}=0.8$. 
Thin and broken lines represent the contributions from 
$\xi_{1{\bf k}}$ and $\xi_{2{\bf k}}$ in eq.(\ref{eq:order-11}).
}
\label{fig:order-6}
\end{figure}
For $j^{-1}= 0.25$, the quasi-particle state is well
defined, which has a visible dispersion in the spectral function.
When the exchange coupling $J_{\perp}$ increases, the 
quasiparticle state becomes less dispersive.
Such flattening effect is more clearly seen in direct plots of 
the dispersion relation shown in Fig. \ref{fig:order-6}.
The dispersive band for smaller $J_\perp$ is mainly contributed
by conduction electrons. On the other hand,
for larger $J_\perp$, conduction electrons and $f$-electrons
are mixed up, which in turn gives rise to the flat
dispersion near the critical point.
In this region, other peaks with large weight 
appear besides the quasiparticle peak in the spectral
function, as seen in Fig. \ref{fig:order-5}. This implies that
 the quasiparticle picture does not hold anymore.

\begin{figure}[htb]
\begin{center}
\includegraphics[width=5.5cm]{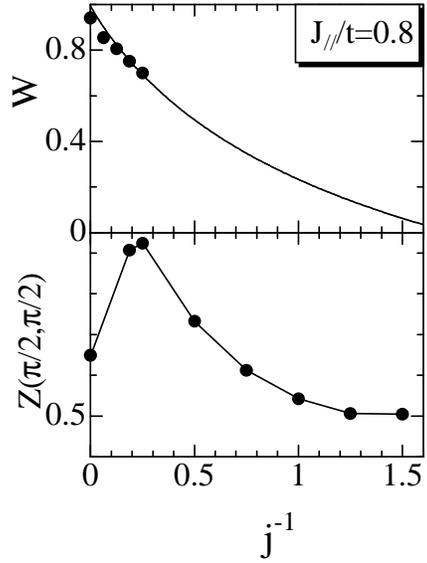}
\end{center}
\caption{
Band 
width $W$ for the quasi-particle and
 renormalization factor $Z$ in the ordered phase.
We show $Z$ with ${\bf k}=(\pi/2, \pi/2)$, since this state has
a rather clear peak structure.
Solid circles are the results for 
the $8\times 8$ system. The solid line for the width 
is computed from the analytic expression eq.(\ref{eq:order-11}).
It is not easy to extract the correct 
values of $W$ from the 
spectral function computed for larger $J_{\perp}$, so that we 
have plotted them only for the small $J_{\perp}$ region.
 The non-monotonic behavior 
in $Z$ reflects a singular 
property of the limit $J_\perp \rightarrow 0$.
}
\label{fig:order-n}
\end{figure}

In Fig. \ref{fig:order-n} we show 
the width of the quasi-particle band 
and the renormalization factor. It is seen that both of these quantities 
become small as the system approaches the critical point, similarly to the 
case of the spin liquid phase discussed in the previous section.
However, in closer comparison of these two cases, we see that 
 the quasi-particle is less stable in the ordered phase, because 
 the quasi-particle state in this case is dressed by gapless
magnon excitations, making it somewhat difficult to
distinguish the quasi-particle state from low-energy
excited continuum.

\section{Summary}\label{Summary}

We have investigated dynamical properties of a single hole doped into 
the Heisenberg-Kondo  model on the square lattice.
At half filling, the model is either in 
the disordered spin-liquid phase or the antiferromagnetically ordered 
phase. In the spin liquid phase, by employing the 
bond-operator representation, 
we have shown that a doped hole has the nature of
a well-defined quasi-particle with 
the unambiguous dispersion relation.
In the vicinity of the quantum critical point, however,
the quasi-particle state is obscured due to the decrease of 
the spin gap.  This effect appears most prominently in the 
region around ${\bf k}=(0,0)$.
Even in this case, it has shown that the 
spectrum around ${\bf k}=(\pi,\pi)$
still forms a sharp peak structure.
In the antiferromagnetically ordered phase, we have encountered
 similar quasi-particle behavior of a doped hole, as in the 
spin-liquid phase. Namely, the quasi-particle state is 
smeared and its dispersion becomes flat as the system
approaches the quantum critical point.

Although apparent properties are quite analogous in these 
two cases, there is the essential 
difference between them for the origin of the quasi-particle state: 
 the quasi-particle in the spin liquid phase is 
stabilized in the presence of the 
spin gap, while that in the ordered phase is dressed by 
low-energy magnon excitations.  We have indeed seen that
this causes the difference in the stability of the 
quasi-particle state around the critical point.
 
In this paper, we have discussed the hole-doping effects 
 by starting from two extreme limits. In these 
approaches,
it is not easy to deal with the properties precisely just at the 
critical point. It is thus desirable to directly investigate
anomalous properties such as the temperature dependence of the 
susceptibility around the critical regime.  
Also, it remains an interesting open problem what kind of
unusual properties are expected 
when the finite density of holes are doped into 
the system around the critical point, which is now under consideration.


\section*{Acknowledgments}
This work was partly supported by a Grant-in-Aid from the Ministry 
of Education, Science, Sports and Culture of Japan. A part of computations 
was done at the Supercomputer Center at Institute for Solid Physics, 
University of Tokyo and Yukawa Institute Computer Facility.

%



\begin{thebibliography}{99}


\bibitem{review}
A. C. Hewson: The Kondo Problem to Heavy Fermions (Cambridge Univ. 
Press, Cambridge 1993).

\bibitem{Kinsulator}
S. Doniach, Physica B+C {\bf 91B}, 231 (1977).

\bibitem{klattice}
See for a review,
H. Tsunetsugu, M. Sigrist and K. Ueda,
Rev. Mod. Phys. {\bf69}, 809 (1997).

\bibitem{Kuramoto}
Y. Kuramoto and K. Miyake, J. Phys. Soc. Jpn. {\bf 59}, 2831 (1990).

\bibitem{Andraka}
B. Andraka and A. M. Tsvelik, Phys. Rev. Lett. {\bf 67}, 2886 (1991).

\bibitem{Hertz}
J. A. Hertz, Phys. Rev. B {\bf 14}, 1165 (1976).

\bibitem{Millis}
A. J. Millis, Phys. Rev. B {\bf 48}, 7183 (1993).

\bibitem{Moriya}
T. Moriya and T. Takimoto, J. Phys. Soc. Jpn. {\bf 64}, 960 (1995).

\bibitem{Continentino}
M. A. Continentino, Phys. Rev. B {\bf 47}, 11587 (1993).

\bibitem{Schlottmann}
P. Schlottmann, Phys. Rev. B {\bf 59}, 12379 (1999).

\bibitem{Coleman}
P. Coleman, Physica B {\bf 259-261}, 353 (1999).

\bibitem{Si}
Q. Si, J. L. Smith, and K. Ingersent, Int. J. Mod. Phys. B {\bf 13}, 2331 (1999).

\bibitem{Rosch}
A. Rosch, Phys. Rev. Lett. {\bf 82}, 4280 (1999).

\bibitem{Lavagna}
M. Lavagna and C. Pepin, Phys. Rev. B {\bf 62}, 6450 (2000).
				  
				  
\bibitem{CeCu6}
H. von Lo\"hneysen, A. Schro\"oder, M. Sieck and T. Trappmann,
Phys. Rev. Lett. {\bf 72}, 3262 (1994).

\bibitem{CePd2Si2}
N. D. Mathur, F. M. Grosche, S. R. Julian, I. R. Walker, 
D. M. Freye, R. K. W. Haselwimmer and G. G. Lonzarich,
Nature {\bf 394}, 39 (1998).

\bibitem{CeNi2Ge2}
F. Steglich, B. Buschinger, P. Gegenwart, M. Lohmann, R Helfrich, 
C. Langhammer, P. Hellmann, L. Donnevert, S. Thomas, A. Link,
C. Geibel, M. Lang, G. Sparn and W. Assmus,
J. Phys. Condens. Matter {\bf 8}, 9909 (1996).

\bibitem{UPd3}
C. L. Seaman, M. B. Maple, B. W. Lee, S. Ghamaty,
M. S. Torikachvili, J.-S. Kang, L. Z. Liu, J. W. Allen, and D. L. Cox,
Phys. Rev. Lett. {\bf 67}, 2882 (1991).

\bibitem{LaRu2Si2}
S. Kambe, S. Raymond, L.-P. Regnault, J. Flouquet, P. Lejay and P. Haen,
J. Phys. Soc. Jpn. {\bf 65}, 3294 (1996). 

\bibitem{Chakravarty}
S. Chakravarty, B. I. Halperin and D. R. Nelson, 
Phys. Rev. Lett. {\bf 60}, 1057 (1988). 

\bibitem{rink}
S. Schmitt-Rink , C. M. Varma and A. E. Ruckenstein, 
Phys. Rev. Lett. {\bf 60}, 2793 (1988).

\bibitem{horsh}
G. Martinez and P. Horsh, Phys. Rev. B {\bf 44}, 317 (1991).

\bibitem{liu}
Z. Liu and E. Manousakis, Phys. Rev. B {\bf 45}, 2425 (1992).

\bibitem{frank}
F. Magsiglio, A. Ruckenstein, Schumitt-Rink and C. Varma, 
Phys. Rev. B {\bf 43}, 10882 (1991).






\bibitem{Zheng}
W. Zheng and J. Oitmaa, cond-mat/0209305; cond-mat/0209307.

\bibitem{Brenig}
C. Jurecka and W. Brenig, Phys. Rev. B {\bf 64}, 092406 (2001).

\bibitem{Assaad}
F. F. Assaad, 
Phys. Rev. Lett. {\bf 83}, 796 (1999).

\bibitem{Shi}
Z. P. Shi, R. R. P. Singh, M. P. Gelfand and 
Z. Wang, Phys. Rev. B {\bf 51}, 15630 (1995).

\bibitem{Wang}
Z. Wang, X. P. Li and D. H. Lee, 
Physica (Amsterdam) {\bf 199B-200B}, 463 (1995).

\bibitem{Matsushita}
Y. Matsushita, M. P. Gelfand and 
C. Ishii, J. Phys. Soc. Jpn. {\bf 66}, 3648 (1997).

\bibitem{Sachdev}
S. Sachdev and N. Bhatt, 
Phys. Rev. B {\bf 41}, 9323 (1990).

\bibitem{Sigrist}
S. Gopalan, T. M. Rice and M. Sigrist, 
Phys. Rev. B {\bf 49}, 8901 (1994).

\bibitem{Lee}
Y. L. Lee, Y. W. Lee and C.-Y. Mou, 
Phys. Rev. B {\bf 60}, 13418 (1999).



\bibitem{jurecka}
C. Jurecka and W. Brenig, 
Phys. Rev. B {\bf 63}, 094409 (2001).

\bibitem{Gelfand}
M. P. Gelfand and R. R. P. Singh, Adv. Phys. {\bf 49}, 93 (2000).

\bibitem{Sakai}
T. Sakai and M. Takahashi, J. Phys. Soc. Jpn. {\bf 58}, 3131 (1989).




\end{thebibliography}
\end{document}